\documentclass[prl,aps,twocolumn,groupedaddress,superscriptaddress]{revtex4-2}
\usepackage{graphicx}
\usepackage[nointegrals]{wasysym} %
\usepackage[export]{adjustbox}
\usepackage{amsmath,amsfonts,amssymb,latexsym}
\usepackage{hhline}
\usepackage{bm}
\usepackage{verbatim}
\usepackage{enumitem}
\usepackage{mathrsfs}
\usepackage{slashed}
\usepackage{empheq}
\usepackage[normalem]{ulem}
\usepackage{manfnt}
\usepackage{amsthm}
\usepackage{ragged2e}

\makeatletter
\newcommand{\fmarki}{\ensuremath{\dagger}}
\newcommand{\fmarkii}{\ensuremath{\ddagger}}
\newcommand{\fmarkiii}{*}

\def\@fnsymbol#1{{\ifcase#1\or \fmarki\or  \fmarkii\or \fmarkiii\else\@ctrerr\fi}}
\makeatother

\newcommand\bpm            {\begin{pmatrix}}
	\newcommand\epm           {\end{pmatrix}}

\newcommand{\bs}{\bigskip}

\def\app#1#2{%
	\mathrel{%
		\setbox0=\hbox{$#1\sim$}%
		\setbox2=\hbox{%
			\rlap{\hbox{$#1\propto$}}%
			\lower1.1\ht0\box0%
		}%
		\raise0.25\ht2\box2%
	}%
}

\newcommand{\ope}\odot

\newcommand{\bi}{\begin{itemize}}
	\newcommand{\ei}{\end{itemize}}

\newtheorem{theorem}{Theorem}

\theoremstyle{definition}
\newtheorem{definition}{Definition}
\theoremstyle{definition}

\newcommand\bd            {\begin{definition}}
	\newcommand\ed            {\end{definition}}
\newcommand\bt            {\begin{theorem}}
	\newcommand\et            {\end{theorem}}
\newcommand\be            {\begin{equation}}
	\newcommand\ee            {\end{equation}}
\newcommand\ba            {\begin{aligned}}
	\newcommand\ea            {\end{aligned}}
\newcommand\bea{\begin{equation}\begin{aligned}}
		\newcommand\eea{\end{aligned}\end{equation}}

\usepackage{hyperref} 
\hypersetup{final}
\hypersetup{colorlinks, citecolor=red, linkcolor=red, urlcolor=red}

\usepackage[mathscr]{eucal} %

\usepackage{braket}

\begin{document}
 \title{Correlated hopping induced topological order in an atomic mixture}
	\author{Ashirbad Padhan}
	\email{padhanashirbad@gmail.com}
	\affiliation{School of Physical Sciences, National Institute of Science Education and Research, Jatni 752050, India}

        \affiliation{Homi Bhabha National Institute, Training School Complex, Anushaktinagar, Mumbai 400094, India}
        \affiliation{Laboratoire de Physique Th\'{e}orique, Universit\'{e} de Toulouse, CNRS, UPS, France}
	
	\author{Luca Barbiero}
	\email{ luca.barbiero@polito.it}
	\affiliation{Institute for Condensed Matter Physics and Complex Systems, DISAT, Politecnico di Torino, I-10129 Torino, Italy}
	
	\author{Tapan Mishra}
	\email{mishratapan@niser.ac.in}
	
\affiliation{School of Physical Sciences, National Institute of Science Education and Research, Jatni 752050, India}

 \affiliation{Homi Bhabha National Institute, Training School Complex, Anushaktinagar, Mumbai 400094, India}

\date{\today}
\begin{abstract}

The large majority of topological phases in one dimensional many-body systems are known to
inherit from the corresponding single-particle Hamiltonian. In this work, we go beyond this assumption and find a new example of
topological order induced through specific interactions couplings. Specifically, we consider a fermionic mixture where one
component experiences a staggered onsite potential and it is coupled through density dependent hopping interactions to the other fermionic component. Crucially, by varying the sign of the staggered potential, we show that this latter fermionic component can acquire topological properties. Thanks to matrix product state simulations, we prove this result both at the equilibrium by extracting the behavior of correlation functions and in an out-of-equilibrium scheme by employing a Thouless charge pumping. Notably, we further discuss how our results can be probed in quantum simulators made up of ultracold atoms. Our results reveal an important and alternative mechanism that can give rise to topological order.

\end{abstract}
\maketitle
	
\paragraph*{Introduction.}
The search for systems exhibiting topological phases and phase transitions have led several discoveries following the observation of the seminal quantum Hall effect in condensed matter~\cite{Klitzing, Thouless1982,KOHMOTO1985343,fqh1, haldane_1, Hasan_rev}. Significant progress has been made in the past decades to obtain signatures of the topological phases and  to understand the origin and stability of such phases in various systems~\cite{Chiu_rev, Qi_2011_rev, senthil_review, rachel_review, Wu_2016}. Hallmarked by the gapped bulk spectrum,  quantized topological invariants and associated gapless  edge modes, the topological phases have attracted a great deal of attention due to their fundamental and technological importance. This has led to their observations in various systems ranging from condensed matter systems, optical systems, meta-materials to mechanical systems~\cite{Atala2013,Takahashi2016pumping,Lohse2016,Mukherjee2017,Lu2014,ssh_expt_1, ssh_expt_2,Kitagawa2012,Leder2016}. 

One of the simplest systems that exhibits topological properties is the one dimensional tight binding model of free fermions with dimerized hopping, known as the Su-Schrieffer-Heeger (SSH) model~\cite{ssh, ssh2, Asboth2016_ssh}. A suitable choice of hopping dimerization of the fermions stabilizes a topological phase which undergoes a transition to a trivial phase by specifically shaping the hopping dimerization. While the topological order in the SSH model is a feature of the single-particle Hamiltonian, such scenario also persists in presence of interactions~\cite{Hayward2018, Nakagawa2018,Mondal2019,Mondal2020, fleischhauer_2013_prl, wang_2015_sop, Santos2018, Karrasch, Bermudez, Sirker_2014, ssh_bosonization, marques, Jiansheng, Yahyavi_2018, liberto1, liberto2, sjia, Marques_2018, Santos2018,mondal_sshhubbard,Yoshida2018,Nakagawa2018,wang_2015_sop,fan2016, Yoshihito,sergi_2022_sop, Le2020, Montorsi_2022, Hetenyi}. Indeed, the topological properties appearing in most of the many-body systems are known to inherit from the single-particle models and in some cases depend on the specific choice of the inter-particle interactions that indirectly favor the SSH-type phenomena~\cite{suman_v1v2,Parida_2024, harsh_padhan_2024,parida_2025}. In other words, the particles experience an induced hopping dimerization under proper conditions, leading to the topological phase. Now the question is whether such a topological phase and associated phase transition in a many-body system can be established without requiring its inheritance from the single-particle limit or through direct inter-particle interactions.

In this Letter, we tackle this interesting question. Specifically, we propose a system of spinful fermions in a one dimensional optical lattice with asymmetric density dependent (DD) tunneling (i.e., the hopping of one component depends on the density of the other and not vice-versa) in which a topological phase transition can be established. By subjecting one of the components to a staggered onsite potential and allowing the hopping of other component to depend on the density of the former, we show that the latter component acquires topological properties. The onset of the topological phase is strongly dependent on the sign of the staggered potential, resulting in a transition from the topological phase to trivial phase at a point where the staggered onsite potential vanishes. We stress that the topological phase that appears in the system considered here is a pure many-body effect and does not inherit from the single-particle Hamiltonian.

\begin{figure}[t!]
\centering
\includegraphics[width=0.7\linewidth]{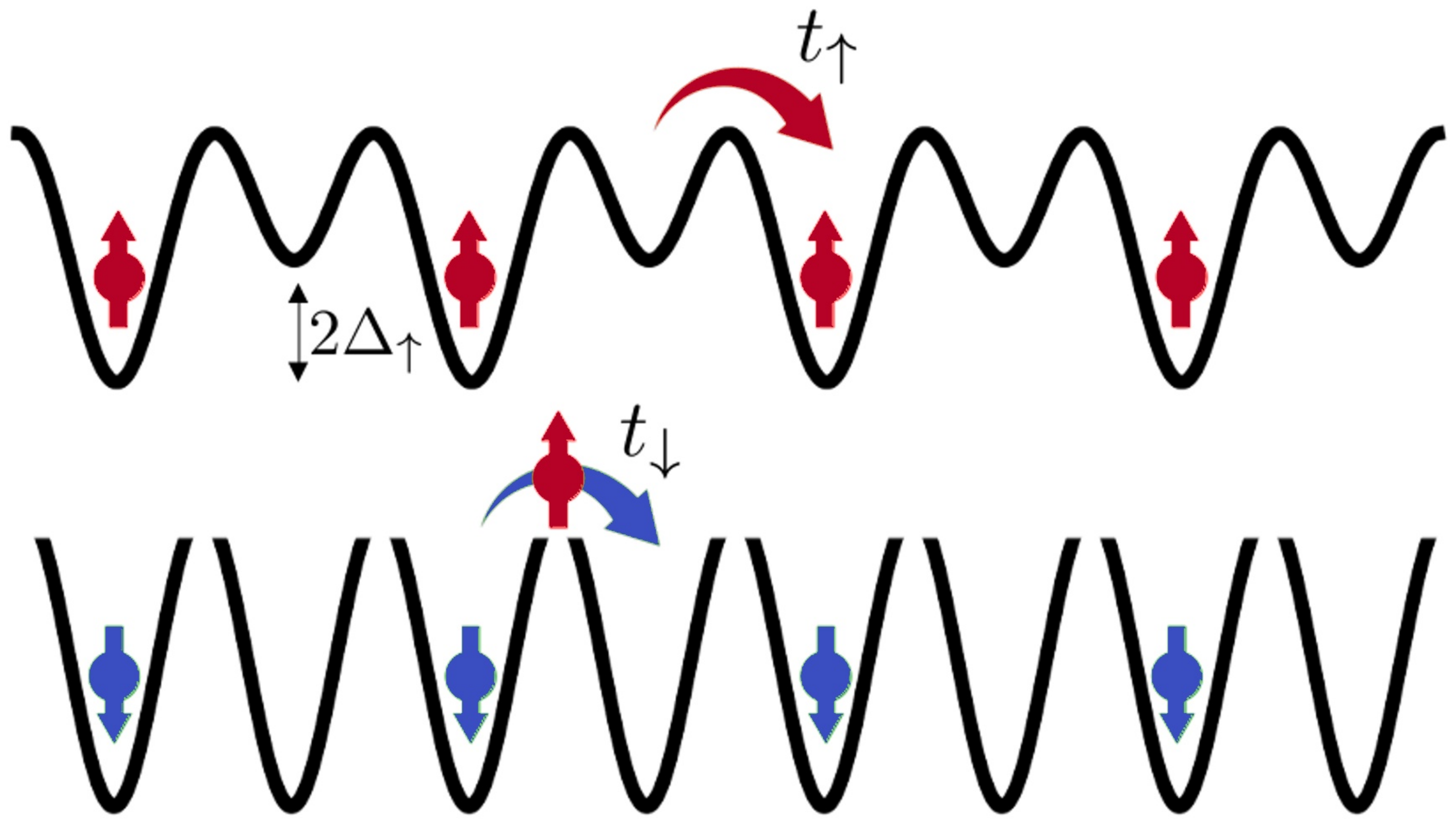}
\caption{Depiction of the model presented in Eq.~\ref{eq:ham} with hopping strengths $t_{\uparrow}$ and $t_{\downarrow}$, and onsite staggered potential strength $\Delta_{\uparrow}$. Here the $\uparrow$-component exhibits a superlattice behaviour thanks to the staggered potential and the hopping of the $\downarrow$-component depends on the onsite density of the $\uparrow$-component.}
\label{fig:lat}
\end{figure}

\paragraph*{Model.} The model of spinful fermions with asymmetric DD tunneling is given by 
\begin{eqnarray}
\label{eq:ham}
    H=-t_{\uparrow}\sum_i(c_{i\uparrow}^{\dagger}c_{i+1\uparrow}+\text{H.c.}) 
    +\Delta_{\uparrow}\sum_i(-1)^in_{i\uparrow}\\ \nonumber 
    -t_{\downarrow}\sum_in_{i\uparrow}(c_{i\downarrow}^{\dagger}c_{i+1\downarrow}+\text{H.c.}),
\end{eqnarray}
where $c_{i\sigma}$ is the fermionic annihilation operator for the  component $\sigma\in(\uparrow,~\downarrow)$. Here, $n_{i\sigma}$ is the respective number operator at the $i^{th}$ lattice site. While $t_{\uparrow}$ and $\Delta_{\uparrow}$ denote the hopping and staggered onsite potential strengths of the $\uparrow$-component, respectively, $t_{\downarrow}$ fixes the hopping strength of the $\downarrow$-component. Note that here the hopping of $\downarrow$-component ($t_\downarrow$) depends on the site occupation of the $\uparrow$-component ($n_{i\uparrow}$), as shown in Fig.~\ref{fig:lat}. The model considered above applies also to a system of two-component hardcore bosons. 

For the model shown in Eq.~\ref{eq:ham}, in the absence of the staggered onsite potential ($\Delta_{\uparrow}=0$) and density dependence of the hopping of the $\downarrow$-component, the system consists of two decoupled gapless Luttinger liquids (LLs) of two individual components even at half filling. In this decoupled LL limit, any finite value of $\Delta_\uparrow$ will turn the $\uparrow$-component gapped, exhibiting a density-wave (DW) type ordering ($\ldots 1~0~1~0 \ldots$) throughout the lattice. Surprisingly, at this point if the hopping of the $\downarrow$-component is made to depend on the density of the $\uparrow$-component through the DD tunneling then the $\downarrow$-component becomes gapped and turns topological if $\Delta_{\uparrow} > 0$. In the following, we discuss this phenomenon and associated topological phase transition in detail. Here, we explore the ground state properties of the model using the matrix product states (MPS) method for a half-filled system, i.e., at density $\rho_{\sigma}=N_{\sigma}/L=1/2$ on a system of $L$ lattice sites and $N_{\sigma}$ number of particles in each component 
We perform all the numerical simulations by considering open boundary conditions and by fixing $t_{\uparrow}=t_{\downarrow}=1$ as the energy unit.

\paragraph*{Results.} 
\begin{figure}[t!]
\centering
\includegraphics[width=1\linewidth]{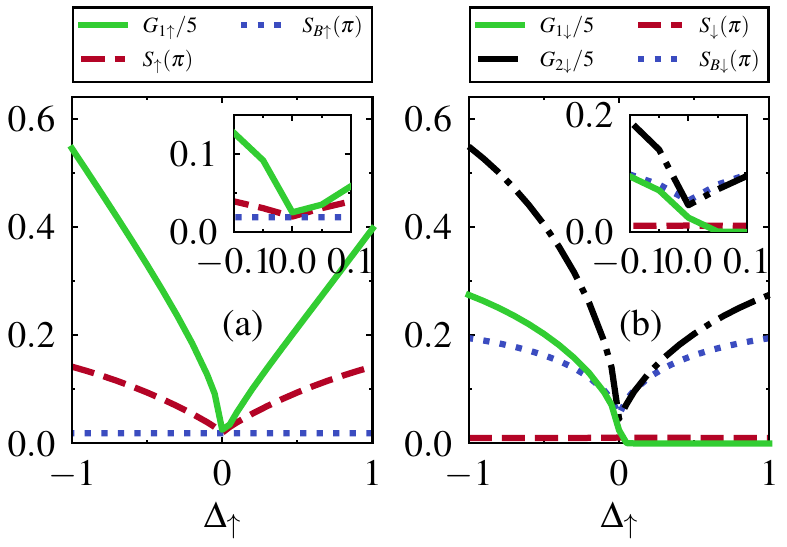}
\caption{(a) Single-particle charge gap $G_{1\uparrow}$ (green solid line), DW structure factor $S_{\uparrow} (\pi)$ (red dashed line) and BO structure factor $S_{B\uparrow} (\pi)$  (blue dotted line) are plotted as a function of $\Delta_{\uparrow}$. (b) Single-particle charge gap $G_{1\downarrow}$ (green solid line), two-particle charge gap $G_{2\downarrow}$ (black dashed-dotted line), DW structure factor $S_{\downarrow} (\pi)$ (red dashed line) and BO structure factor $S_{B\downarrow} (\pi)$  (blue dotted line) are plotted as a function of $\Delta_{\uparrow}$. The insets show a zoomed-in view of the respective figures. Here we extrapolate the values of $G_{1,2\sigma}$ with system sizes $L=40, 60, 80$ and $100$, and consider $L=100$ for the calculation of $S_{\sigma} (\pi)$ and $S_{B\sigma} (\pi)$. The charge gaps are scaled by a factor of $5$ for better clarity. To avoid edge effects, the structure factors are calculated in the bulk of the lattice, i.e., in the range $L/4$ to $3L/4$.}

\label{fig:sshpd}
\end{figure}
We first analyze the equilibrium properties of the system. As already mentioned before, for any finite values of $\Delta_\uparrow$, the $\uparrow$-component is expected to form a gapped phase of DW type ordering at half filling, i.e., at $\rho_{\uparrow}=1/2$. Now, through the DD tunneling $t_\downarrow$, the properties of the $\downarrow$-component are strongly affected by the density of the $\uparrow$-particles. 
First of all, we confirm the gapped DW type phase of the $\uparrow$-component by studying the system's gaps. To quantify them we compute the single- and two-particle excitation gaps (or the charge gaps) of the system which are defined as
\begin{eqnarray}
G_{1\sigma}=E(N_{\sigma}+1)+E(N_{\sigma}-1)- 2E(N_{\sigma})
\end{eqnarray}
and
\begin{eqnarray}
G_{2\sigma}=E(N_{\sigma}+2)+E(N_{\sigma}-2)- 2E(N_{\sigma}),
\end{eqnarray}
respectively. Here, $E(N_{\sigma})$ is the ground state energy with $N_{\sigma}$ number of particles in each component. In Fig.~\ref{fig:sshpd}(a), we plot $G_{1\uparrow}$ as a function of $\Delta_\uparrow$ (green solid line). It can be seen that for all values of $\Delta_\uparrow$, the system becomes gapped even in the limit of $\Delta_{\uparrow}$ becoming zero (see inset of Fig.~\ref{fig:sshpd}(a)). Due to the staggered onsite potential, the particles are expected to arrange themselves in a DW pattern due to doubling up of the unit cell. This behaviour can be understood from the structure factor defined as 
\begin{eqnarray}
\label{eq:cdwstfc}
S_{\sigma}(k) = \frac{1}{L^2}\sum_{i,j}e^{\iota k|i-j|}\langle n_{i\sigma}n_{j\sigma}\rangle.
\end{eqnarray}
We plot $S_{\uparrow}(\pi)$ as a function $\Delta_\uparrow$ (red dashed line) in Fig.~\ref{fig:sshpd}(a) which becomes finite as soon as the system becomes a gapped DW for any finite values of $\Delta_\uparrow$. Note that the finite structure factor is not a result of spontaneous symmetry breaking in the system rather due to the staggered onsite potential $\Delta_\uparrow$. Now, we will show that this DW ordering in the $\uparrow$-component is the key to induce topological features in the initially non-topological $\downarrow$-component through the DD tunneling. 

For this purpose, we first monitor the behaviour of the gap $G_{1\downarrow}$ as a function of $\Delta_\uparrow$ which is shown in Fig.~\ref{fig:sshpd}(b). Contrary to the $\uparrow$-component case, it can be seen that $G_{1\downarrow}$ (green solid line) remains finite in the regime $\Delta_\uparrow \leq 0$ and vanishes when $\Delta_\uparrow > 0$. Such vanishing up of the single-particle gap in the regime $\Delta_\uparrow > 0$ does not mean that the system is gapless rather it is an indication of the existing edge modes within the gap at half filling. In such a situation the gap in the system can be monitored by the two particle gap $G_{2\downarrow}$ (black dashed-dotted line) as shown in  Fig.~\ref{fig:sshpd}(b) which remains finite even for $\Delta_\uparrow > 0$. This behaviour of gap indicates a gapped-gapped transition as a function of $\Delta_{\uparrow}$ through $\Delta_\uparrow=0$.

We find that while the gaps in the $\uparrow$-component correspond to a DW type ordering in the system, for the $\downarrow$-component they represent the dimerized or the bond-order (BO) phases. We prove the BO nature of the phase from the BO structure factor 
\begin{eqnarray}
\label{eq:bostfc}
    S_{B\sigma}(k) = \frac{1}{L^2}\sum_{i,j}e^{\iota k|i-j|}\langle B_{i\sigma}B_{j\sigma}\rangle,
\end{eqnarray} 
where $B_{i\sigma}= (c_{i\sigma}^{\dagger}c_{i+1 \sigma}+\text{H.c.})$ is the bond energy associated to the $i^{th}$ bond of the component $\sigma$.
In Fig.~\ref{fig:sshpd}(b), we plot  $S_{B\downarrow}(\pi)$ (blue dotted line) as a function of $\Delta_{\downarrow}$ which exhibits finite values for $|\Delta_\uparrow| \neq 0$ indicating the BO nature of the gapped phases. To rule out the DW type  ordering as exhibited by the $\uparrow$-particles, we also plot $S_{\downarrow}(\pi)$ (red dashed line) which vanishes for all the values of $\Delta_\uparrow$. This confirms that the gapped phases for $|\Delta_\uparrow| \neq 0$ are the BO phases. Note that such bond ordering is absent for the $\uparrow$-particles which can be seen from the vanishing values of $S_{B\uparrow}(\pi)$ (blue dotted line) in Fig.~\ref{fig:sshpd}(a).  

The above analysis clearly shows that the $\downarrow$-component develops bond ordering due to the DW type structure in the $\uparrow$-component through the DD tunneling. Now the question is how the two BO phases on either sides of $\Delta_{\uparrow}=0$ in Fig.~\ref{fig:sshpd}(b) are related to each other. In the following we will show that while the BO phase  for $\Delta_\uparrow > 0$ exhibits  topological characters, for $\Delta_\uparrow < 0$ it is a trivial BO phase, making the BO phases topologically distinct from each other.

\paragraph*{Topological properties.} 

\begin{figure}[t!]
\centering
\includegraphics[width=1\linewidth]{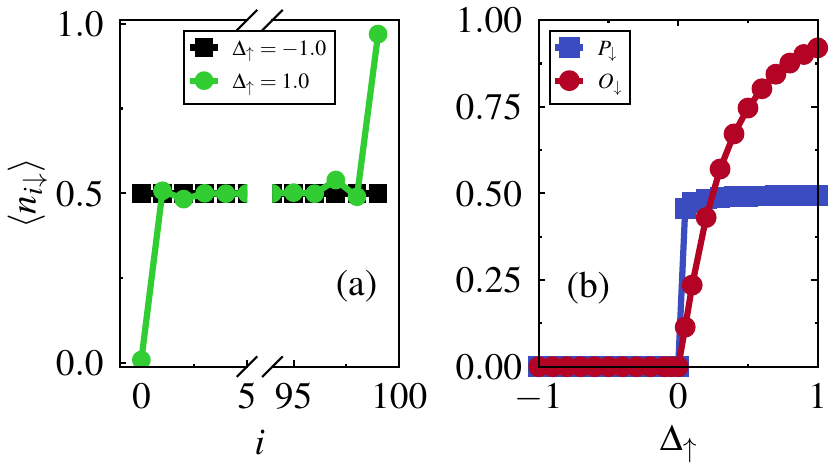}
\caption{(a) Onsite densities $\langle n_{i\downarrow}\rangle$ are plotted as a function of site index $i$ for $\Delta_{\uparrow}=-1.0$ (black line with squares) and for $\Delta_{\uparrow}=1.0$ (green line with circles). (b) Polarization $P_{\downarrow}$ (blue line with squares) and string order parameter $O_{\downarrow}$ (red line with circles) are plotted as a function of $\Delta_{\uparrow}$. Here the system size considered is $L=100$.}
\label{fig:ops}
\end{figure}

First of all, we characterise the topological phase from the appearance of the edge states. As depicted in Fig.~\ref{fig:ops}(a), the presence of degenerate edge states appear rather clearly by extracting the onsite densities $\langle n_{i\downarrow}\rangle$ (green circles) for positive values of $\Delta_{\uparrow}$. On the other hand, for negative values of $\Delta_{\uparrow}$, the system does not develop any edge states as can be seen from the values of $\langle n_{i\downarrow}\rangle \sim 0.5$ (black squares) in Fig.~\ref{fig:ops}(a). 
To quantify such topological character we compute 
the edge state polarization defined by the formula
\begin{eqnarray}
\label{eq:pol}
P_{\sigma}=\frac{1}{L}\sum_i(i-i_0)\langle\psi|n_{i\sigma}|\psi\rangle,
\end{eqnarray}
where $\ket{\psi}$ is the ground state wavefunction and $i_0=(L-1)/2$ is the centre-of-mass position of the lattice~\cite{Mondal2020}. This quantity is directly accessible in experiments through measuring the centre-of-mass shift of the particle densities~\cite{Lohse_2015, Takahashi2016pumping,Citro_2023, Walter2023}. In Fig.~\ref{fig:ops}(b), we plot $P_{\downarrow}$ as a function of $\Delta_{\uparrow}$ which shows that the polarization clearly vanishes at the critical point $\Delta_\uparrow =0$ as well as in the BO phase for $\Delta_\uparrow < 0$. However, for $\Delta_\uparrow > 0$, the polarization assumes a constant value of $P_{\downarrow}\sim0.5$.  The finite value of polarization is usually attained when one of the edge lattice sites is filled and the other is not. However, when the particle density is uniformly distributed along the lattice, it becomes zero. Thus, $P_{\downarrow}$ together with the onsite density $\langle n_{i\downarrow}\rangle$ indirectly checks the existence of non-trivial edge states in a system.  This indicates that the BO phase in the regime of $\Delta_\uparrow > 0$ is topological in nature. The topological BO phase is further quantified by the non-local string order parameter defined by 
\begin{eqnarray}
\label{eq:ostr}
O_{\sigma}=-\langle z_{i\sigma}e^{\iota\frac{\pi}{2}\sum_{k=i+1}^{j-1}z_{k\sigma}}z_{j\sigma}\rangle,
\end{eqnarray}
where $z_{i\sigma} = 1-2n_{i\sigma}$~\cite{fraxanet_2021_sop, wang_2015_sop, sergi_2022_sop, Anfuso_2007_sop, Endres_2011_sop, Haller_2015_sop, Hilker_2017_sop, Sylvain_2019_sop, Mazurenko_2017_sop, Montorsi_2022}. Here we avoid the edge lattice sites and consider the maximum possible distance by choosing $i=2$ and $j=L-1$.
In our case $O_{\downarrow}$ (red circles) becomes finite in the topological BO phase (for $\Delta_\uparrow > 0$) but vanishes elsewhere as shown in Fig.~\ref{fig:ops} (b). 

These findings reveal that while the DD tunneling turns the $\downarrow$-component into a BO phase, depending on the sign of $\Delta_\uparrow$, the BO phase can be either trivial ($\Delta_\uparrow < 0$) or topological ($\Delta_\uparrow > 0$) in nature and hence a trivial to topological phase transition occurs as a function of $\Delta_\uparrow$ through the critical point at $\Delta_\uparrow = 0$. Such a non-trivial transformation of a non-topological phase to a topological phase is due to the interplay of the onsite staggered potential and the DD tunneling which can be understood from the following arguments.

\begin{figure}[t!]
\centering
\includegraphics[width=1\linewidth]{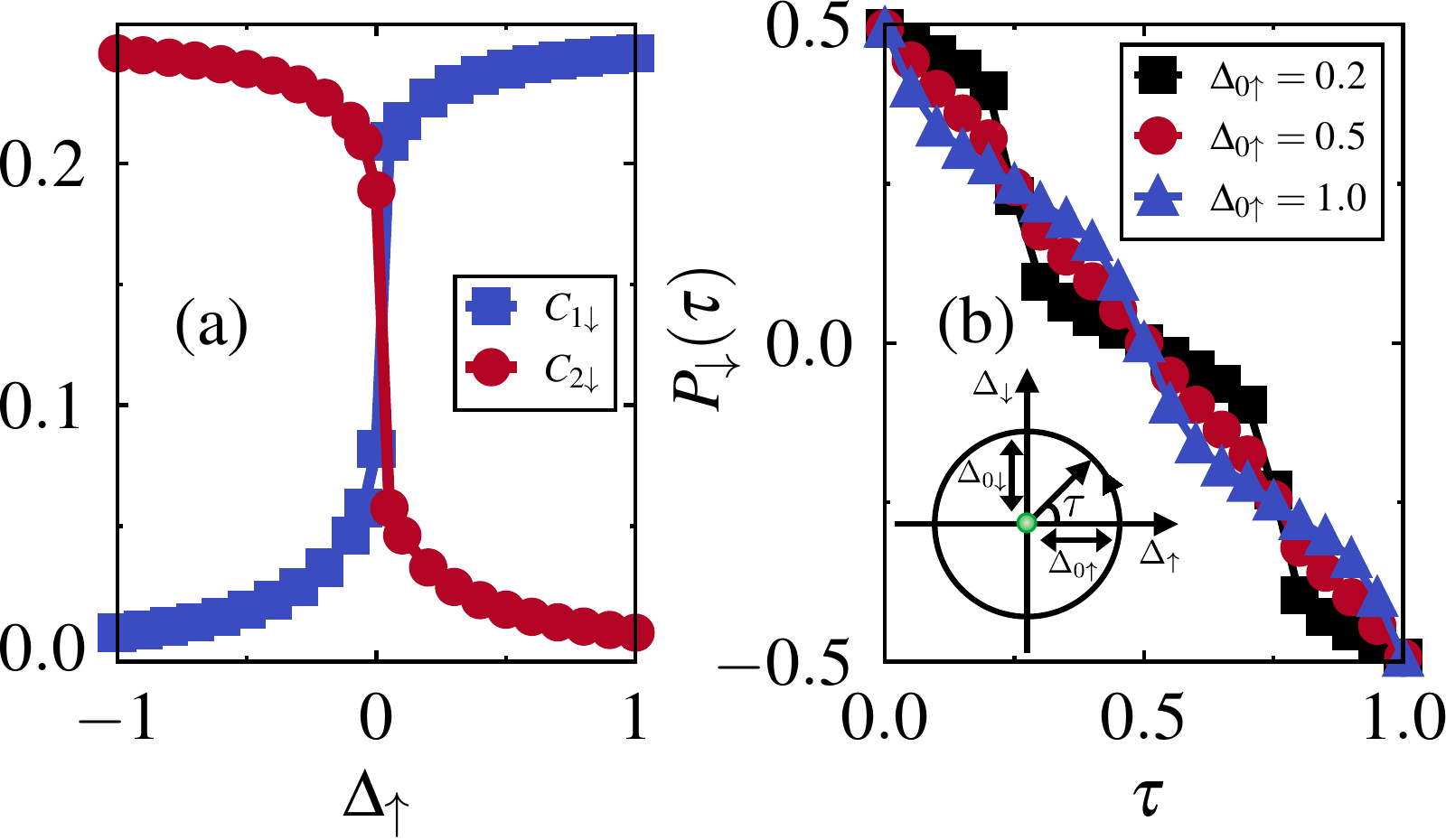}
\caption{(a) Intra- and inter-cell density-density correlations $C_{1\downarrow}$ (blue line with squares) and $C_{2\downarrow}$ (red line with circles), respectively, are plotted as functions of $\Delta_\uparrow$. To avoid edge effects, $C_{1\downarrow}$ and $C_{2\downarrow}$ are calculated in the bulk of the lattice, i.e., in the range $L/4$ to $3L/4$. (b) Polarization $P_{\downarrow}$ is plotted as a function of the pumping parameter $\tau$ for $\Delta_{0\uparrow}=0.2$ (black line with squares), $0.5$ (red line with circles) and $1.0$ (blue line with triangles). Here, we fix $\Delta_{0\downarrow}=0.5$ and consider a system of size $L=100$. The inset shows the pumping scheme in the $\Delta_{\uparrow}$-$\Delta_{\downarrow}$ plane where the green dot marks the critical point, i.e., $(\Delta_{\uparrow}, \Delta_{\downarrow})=(0, 0)$.}
\label{fig:pump}
\end{figure}

Due to the difference in the chemical potentials (or onsite energies) on alternate lattice sites, the $\uparrow$-particles get trapped in the deeper lattice sites, i.e., the sites with negative potentials. On the other hand, the hopping along the $i^{th}$ bond of the $\downarrow$-component becomes finite only if the density at the $i^{th}$ lattice site of the $\uparrow$-component is finite. Thus, disconnected dimers of $\downarrow$-component are formed resulting in a fully dimerized effective SSH model. Therefore, depending on whether the first site is empty or filled for the $\uparrow$-component due to the choice of $\Delta_\uparrow$, the $\downarrow$-component becomes topological or trivial in nature. This picture can be clearly understood from the average nearest-neighbour inter- and intra-cell density-density correlation functions defined as
\begin{eqnarray}
\label{eq:corr1}
    C_{1\sigma} = \frac{1}{L}\sum_{i\in\text{even}}\langle n_{i \sigma} n_{i+1 \sigma}\rangle
\end{eqnarray}
and
\begin{eqnarray}
\label{eq:corr2}
    C_{2\sigma} = \frac{1}{L}\sum_{i\in\text{odd}}\langle n_{i \sigma} n_{i+1 \sigma}\rangle,
\end{eqnarray}
respectively. Since the inter-cell (intra-cell) correlation is dominant in the topological (trivial) BO phase, $C_{1\downarrow}$ ($C_{2\downarrow}$) is expected to become finite. From Fig.~\ref{fig:pump}(a) we can notice that $C_{1\downarrow}$ (blue squares) remains very small for all the values of $\Delta_\uparrow < 0$. After the critical point, i.e., for $\Delta_\uparrow >0$, there is a sharp increase in $C_{1\downarrow}$ indicating the topological BO phase. At the same time $C_{2\downarrow}$ (red circles) behaves just the opposite way to $C_{1\downarrow}$.

After discussing the equilibrium scenarios, in the remaining part of the paper we explore the signatures of the topological phase transition through Thouless charge pumping.

\paragraph*{Thouless pumping.} Thouless charge pumping or the topological charge pumping (TCP), deals with the quantized transport of particles in a topological system through an adiabatic periodic modulation of system parameters. Moreover, they can not only probe the topological phase transitions but also provide a unique platform for a possible experimental probing~\cite{Takahashi2016pumping, Spielman, bloch_spinpump, Sylvain, Esslinger_expt, Wang_expt, Zilberberg_expt, J_rgensen_2021, J_rgensen_2023, Cheng_2022, Liu_expt, Ke_2016, Cheng_expt, Grinberg2020, tao2023interactioninduced, viebahn2023interactioninduced, Hatsugai_2016, Wang, Chaohong_2020, Mondal2020, suman_v1v2, Kuno_int, Nakagawa2018, bertok_pump, mondal_phonon, meden, Hayward2018, mondal_sshhubbard, Barbiero, Cooper_rev, Ozawa_rev, hayward, Takahashi,tatp, floquet1d, HOTI, Cerjan2020, Padhan_pumping_2024}. 
In the following, we propose a pumping scheme to capture the topological phase transition of the $\downarrow$-component described in the previous section. We define the pumping Hamiltonian in the spirit of the Rice-Mele model~\cite{rm, Asboth2016_rm} as
\begin{eqnarray}
\label{eq:hamp}
    H_{p}=-t_{\uparrow}\sum_i(c_{i\uparrow}^{\dagger}c_{i+1\uparrow}+\text{H.c.}) 
    +\Delta_{\uparrow}\sum_i(-1)^in_{i\uparrow}\\ \nonumber 
    -t_{\downarrow}\sum_in_{i\uparrow}(c_{i\downarrow}^{\dagger}c_{i+1\downarrow}+\text{H.c.}) + \Delta_{\downarrow}\sum_i(-1)^in_{i\downarrow},
\end{eqnarray}
where the first three terms are same as Eq.~\ref{eq:ham} and the last term is an extra staggered onsite potential term associated to the $\downarrow$-component. The periodic modulation of the onsite potentials are achieved as $\Delta_{\uparrow}=\Delta_{0\uparrow}\cos(2\pi\tau)$ and $\Delta_{\downarrow}=\Delta_{0\downarrow}\sin(2\pi\tau)$. When the pumping parameter $\tau$ which is equivalent to time is varied from $0$ to $1$, a closed path in the parameter space is achieved as shown in the inset of Fig.~\ref{fig:pump}(b). To this end we propose a pumping scheme involving these two onsite terms and consider three pumping cycles in the $\Delta_{\uparrow}$-$\Delta_{\downarrow}$ plane so that the path encloses the critical point, ($\Delta_{\uparrow},\Delta_{\downarrow})=(0, 0)$. The three cycles have the parameters $\Delta_{0\uparrow}=0.2, 0.5$ and $1.0$, respectively, and $\Delta_{0\uparrow}=0.5$ for all the cycles.

We characterize the TCP through the edge polarization using Eq.~\ref{eq:pol} where the ground state $\ket{\psi(\tau)}$ is computed as a function of $\tau$. The polarization is related to the total amount of pumped charge per cycle as
    $Q_{\sigma} = \int_{0}^{1}d\tau\partial_{\tau}P_{\sigma}(\tau)$,
which takes quantized values if all the criteria for a robust pumping are satisfied. In Fig.~\ref{fig:pump}(b) we plot $P_{\downarrow}(\tau)$ for $\Delta_{0\uparrow}=0.2$, which shows a smooth variation from $0.5$ to $-0.5$, thus, a total pumped charge of $|Q_{\downarrow}|=1$. Similarly, for both $\Delta_{0\uparrow}=0.5$ and $\Delta_{0\uparrow}=1.0$, the polarization varies smoothly, which signifies a quantized and robust charge pumping. This clearly indicates that the $\downarrow$-component exhibits a topological to trivial transition as $\Delta_\uparrow$ varies from negative to positive values through the critical point at $\Delta_\uparrow=0$.

\paragraph*{Conclusions.} We have proposed a model of spinful fermions in one dimension where a topological phase can be stabilised through the correlated tunneling.  By imposing staggered onsite potential on one of the components, we have shown that a topological phase can be induced in the other component if the hopping of the latter component depends on the onsite density of the former component. Such induced topological phase in turn results in a phase transition from a trivial BO phase to topological BO phase  as a function of the onsite staggered potential. 
While the topological and trivial phases exhibit characters similar to the ones exhibited by the SSH model, the phase transition does not occur through a gap closing point. Moreover, the bond ordering in this case arises from the competing effects of the onsite potential and the density-dependent tunneling as opposed to the dimerized hopping in the SSH model.

It is crucial to underline that, while on one hand, our results unveil a novel mechanism to generate topological phases, on the other, our results can be of crucial relevance also from an experimental perspective. It is indeed important to stress that, contrary to previous proposals \cite{sergi_2022_sop}, our schemes works for finite systems in presence of boundaries. This represents an important aspect as this is the usual configuration in quantum simulators made up of ultracold atoms in optical lattices which can allow both for the experimental realization of the model in Eq.~\ref{eq:ham} and for an efficient probing of its phase diagram. More in details, resonant Floquet drivings of interacting atoms \cite{Kolovsky_2011,Chen2011,Goldman2015} combined with magnetic field gradients or superlattice potentials have already paved the way towards the experimental realization of systems where the hopping of only one component is affected by the occupation of the other one \cite{Bermudez_2015,Barbiero2019,Schweizer2019,Gorg2019}. In addition, staggered onsite potentials are largely implemented in ultracold atoms' experiments \cite{Lohse2016,Takahashi2016pumping,Esslinger_expt,Walter2023}, thus making evident that the model in Eq.~\ref{eq:ham} might be promptly realized. Finally, through quantum gas microscopy \cite{Gross2021} the onsite density can be measured with high accuracy. In such  a way, the behavior of both the edge state polarization (Eq.~\ref{eq:pol}) and the string correlator (Eq.~\ref{eq:ostr}) can be probed in order to provide an accurate characterization of the topological properties of the system.

\paragraph*{Acknowledgments.} We thank Subhro Bhattacharjee for useful discussion. T.M. acknowledges support from Science and Engineering Research Board (SERB), Govt. of India, through project No. MTR/2022/000382 and STR/2022/000023. L. B. acknowledges financial support within the DiQut Grant No.
2022523NA7 funded by European Union – Next
Generation EU, PRIN 2022 program (D.D. 104
- 02/02/2022 Ministero dell’Università e della
Ricerca).

\bibliography{references}

\bs

\end{document}